\documentclass{article}

\usepackage[english]{babel}
\usepackage[letterpaper,top=2cm,bottom=2cm,left=3cm,right=3cm,marginparwidth=1.75cm]{geometry}
\usepackage[square,numbers,sort&compress]{natbib}
\usepackage[font=small,labelfont=bf]{caption}

\usepackage{amsmath}
\usepackage{graphicx}
\usepackage[colorlinks=true, allcolors=blue]{hyperref}
\usepackage{xcolor}

\usepackage{color-edits}
\addauthor{JWV}{red}

\addauthor{VL}{blue}

\title{AI Transparency in the Age of LLMs: A Human-Centered Research Roadmap}
\author{Q. Vera Liao and Jennifer Wortman Vaughan\footnote{equal contribution}\\ Microsoft Research}

\begin{document}
\maketitle

\begin{abstract}
The rise of powerful large language models (LLMs) brings about tremendous opportunities for innovation but also looming risks for individuals and society at large. We have reached a pivotal moment for ensuring that LLMs and LLM-infused applications are developed and deployed responsibly. However, a central pillar of responsible AI---transparency---is largely missing from the current discourse around LLMs. It is paramount to pursue new approaches to provide transparency for LLMs, and years of research at the intersection of AI and human-computer interaction (HCI) highlight that we must do so with a human-centered perspective: Transparency is fundamentally about supporting appropriate human understanding, and this understanding is sought by different stakeholders with different goals in different contexts. In this new era of LLMs, we must develop and design approaches to transparency by considering the needs of stakeholders in the emerging LLM ecosystem, the novel types of LLM-infused applications being built, and the new usage patterns and challenges around LLMs, all while building on lessons learned about how people process, interact with, and make use of information. We reflect on the unique challenges that arise in providing transparency for LLMs, along with lessons learned from HCI and responsible AI research that has taken a human-centered perspective on AI transparency. We then lay out four common approaches that the community has taken to achieve transparency---model reporting, publishing evaluation results, providing explanations, and communicating uncertainty---and call out open questions around how these approaches may or may not be applied to LLMs. We hope this provides a starting point for discussion and a useful roadmap for future research.
\end{abstract}

\section{Introduction}

Hugely powerful large language models (LLMs) like GPT-4, LaMDA, and LLaMA are now being deployed in applications from search engines to code generation tools to productivity suites. These generative models are widely expected to have impact across industries, changing the way we engage in tasks like writing, programming, and design, and reshaping occupations in medicine, law, marketing, education, and beyond~\cite{AGG22,bommasani2021opportunities,LFK23,FRS23,DM22,DL23}. As the chair of the U.S. Federal Trade Commission put it in a recent op-ed, ``the full extent of generative AI's potential is still up for debate, but there's little doubt it will be highly disruptive''~\cite{ftcoped2023}.

While the capabilities of LLMs are impressive, they also raise new risks~\cite{GPT4tech,bender2021stochastic,WMR+22,KBN+23}.  Language models are found to encode biases~\cite{rae2021scaling,abid2021persistent}, which risks propagating harmful discrimination, stereotypes, and exclusion at scale. They are widely known to ``hallucinate'' information~\cite{hallucinating2022,bender2021stochastic,LHE22,MNBM20}, producing outputs that are plausible---even convincing---but incorrect. They may project confidence about these hallucinated outputs, potentially contributing to automation bias, overreliance, or automation-induced complacency~\cite{parasuraman2010complacency, wickens2015complacency}. LLMs can generate harmful, sometimes toxic content, including hate speech and offensive language~\cite{bender2021stochastic,weidinger2021ethical}, or reveal sensitive information that threatens privacy or security.  They can contribute---both intentionally and unintentionally---to the spread of misinformation~\cite{ZZL+23,KMB22,BLMS21}. And in the longer term, LLMs may lead to environmental harms~\cite{bender2021stochastic} as well as socioeconomic harms, including the displacement and deskilling of workers across industries~\cite{weidinger2021ethical}.

Given the anticipated impact that LLMs will have on both our day-to-day lives and society at large, it is critical that LLMs and LLM-infused applications be developed and deployed responsibly. One central component of responsible AI development and deployment is \emph{transparency}: enabling relevant stakeholders to form an appropriate understanding of a model or system's capabilities, limitations, how it works, and how to use or control its outputs.  Developers of LLMs cannot debug their models, responsibly assess whether they are ready to launch, and enforce responsible and safe usage policies for their models without some understanding of their behavior and performance on different tasks.  Business decision-makers, designers, and developers building LLM-infused applications must be able to understand the LLM's capabilities and limitations in order to ideate and make decisions about whether, where, and how to use the model---potentially including how to fine-tune, prompt, or otherwise adapt the model to better fit their use case.  End-users must be able to form a sufficiently accurate understanding of LLM-infused applications to control the application's behavior and achieve appropriate levels of trust and reliance. People impacted by LLMs or LLM-infused applications may require transparency in order to understand their options for recourse. Additionally, given the speed at which powerful new LLMs and their applications are being released and the growing concerns over potential harms, we should expect to see an increased demand for transparency around their development and inner workings from policymakers and third-party auditors aiming to regulate and oversee their use.

In recent years, we have witnessed the creation of a whole research field at the intersection of AI and human-computer interaction (HCI) that is focused on developing and evaluating different approaches to achieve transparency. These approaches range from frameworks for documenting models and the datasets they are trained on~\cite[e.g.,][]{mitchell2019model,crisan2022interactive,arnold2019factsheets,gebru2021datasheets,BF18,holland2018dataset}
to techniques for producing explanations of individual model outputs~\cite[e.g.,][]{RSG16,LL17,R19b,USL19,KL17}
to approaches for communicating uncertainty~\cite[e.g.,][]{bhatt2021uncertainty,dhami2022,wang2021} and beyond. There is no one-size-fits-all solution. In the case of LLMs, the needs of an application developer engaging in ideation are probably different from those of a writer who is using an LLM-infused application to edit a novel or a public figure who is concerned about how their life is presented by an LLM-infused search engine. In our own work~\cite{liao2021human,vaughan2021humancentered}, we have argued for the importance of taking a human-centered perspective on transparency---designing and evaluating transparency approaches with stakeholders and their goals in mind. We believe that this is even more important in the era of LLMs, when the diversity of stakeholders and their experience levels, contexts, goals, and transparency needs, is greater than ever. 

In this paper, we map out a human-centered research roadmap for transparency in this new era. We first reflect on the unique challenges that arise in providing transparency for LLMs compared with smaller-scale, more specialized models that have traditionally been the focus of AI transparency research.  We reflect on lessons learned from HCI and Responsible AI/FATE (fairness, accountability, transparency, and ethics) research that centers on human needs of, interactions with, and impact from AI transparency.  We then lay out common approaches, including techniques and artifacts, that the community has taken to achieve transparency and call out open questions around how they may or may not be applied to LLMs.

We note that there is no agreed-upon definition of transparency, and indeed, transparency has been recognized as a multi-faceted concept. In this paper, we adopt a focus on \textit{informational} transparency---essentially, what information about a model (or system building on that model) should be disclosed to enable appropriate understanding---which has been emphasized within the machine learning (ML) research community and in industry practice, though we note that there are other perspectives, such as the normative, relational, and social dimensions of transparency, that have been studied in the broader literature~\cite{felzmann2020towards,meijer2013understanding}. Some of the approaches we cover, such as model reporting, are primarily aimed at supporting a \textit{functional} understanding of \textit{what} the model (or system) can do, often by exposing the goals, functions, overall capabilities, and limitations. Others, like the explanations frequently explored in the explainable AI (XAI) and interpretable ML communities, are primarily aimed at supporting a \textit{mechanistic} understanding of \textit{how} the model (or system) works, by disclosing the parts and processes ~\cite{lombrozo2012explanation}. We believe that both understandings play important roles and the appropriate form of transparency in any given context will depend on the stakeholder and the goal that they wish to achieve. 

Finally, we note that many of the challenges, lessons learned, potential approaches, and open problems that we call out in this paper apply not only to LLMs but to other large-scale generative models, including multimodal models that allow for both textual and visual input or output. While we adopt the narrower focus on LLMs for simplicity, we encourage additional research on transparency for these other models.

\section{What Makes Transparency for LLMs Challenging?}
\label{sec:challenges}

To ground the discussion in the remainder of the paper, we first explore the unique characteristics of LLMs and the emerging patterns of their usage that are likely to make it more challenging to achieve transparency compared with the smaller-scale, specialized models that AI transparency research has traditionally dealt with. We start by providing some brief background on LLMs and establishing some terminology that we will use in the rest of the paper.

\subsection{Background on LLMs}

An LLM, like any language model, predicts the conditional probability of a token---which might be a character, word, or other string---given its preceding context and, in the case of bidirectional models, its surrounding context~\cite{bengio2003neural,radford2019gpt2}. Present-day LLMs are based on modern neural network self-attention architectures like the transformer~\cite{vaswani2017attention} with hundreds of billions or even more than a trillion parameters~\cite{ganguli2022predictability}. While earlier models were trained on datasets of moderate size, LLMs are trained on datasets of massive scale, with hundreds of billions or even more than a trillion tokens~\cite{HBM+22,Borgeaud2021ImprovingLM}, requiring many orders of magnitude more compute time. This makes LLMs vastly more sophisticated and expressive than their predecessors.

While a basic pre-trained LLM model can be viewed as a ``general-purpose'' next-word predictor, LLMs can be adapted to exhibit or suppress specific behaviors or to perform better on specific tasks like text summarization, question answering, or code generation. One common approach is fine-tuning, in which the model's parameters are updated based on additional, specialized data~\cite[e.g.,][]{howard-ruder-2018-universal,LCK20,devlin-etal-2019-bert,RNSS18}.  A popular technique for fine-tuning is reinforcement learning from human feedback (RLHF) in which human preferences are used as a reward signal~\cite{CLB+17,ouyang2022training}. Another approach is prompting or prompt engineering, in which natural-language prompts---often containing examples of tasks (for few-shot prompting/in-context learning) or demonstrations of reasoning (for chain-of-thought prompting)---are provided to the model to alter its behavior without making any changes to the model's internal parameters~\cite[e.g.,][]{brown2020language,wei2022chain,liu-etal-2022-makes,shin-etal-2020-autoprompt}. The adapted model can be incorporated into applications such as chatbots, search engines, or productivity tools. Models can also be augmented with the ability to call on external models, tools, or plugins~\cite{mialon2023augmented}, for example, querying an information retrieval system to ground their output or controlling and receiving feedback from a physical robot. 

It is important to note that the party adapting the model or building the application is frequently not the same party who built the underlying pre-trained LLM, and may only be able to access the LLM through an API. A model may also be adapted more than once by different parties; for instance, a base model may be fine-tuned using RLHF by its creators, fine-tuned on domain-specific data by application developers, and then adapted via in-context learning by end-users.  When we talk about transparency, we must keep in mind whether we are referring to transparency about the pre-trained LLM, an adapted LLM, or the application using the pre-trained or adapted model (LLM-infused application). We aim to call out which of these we are referring to when it is not clear from the context.

\subsection{Challenges for Achieving Transparency}

There are several characteristics of LLMs and their usage that pose challenges for transparency. The list we lay out here is not meant to be exhaustive, but to provide context for later discussion.

\paragraph{Complex and Uncertain Model Capabilities and Behaviors.} 

LLMs can perform an astonishingly wide variety of tasks in different contexts~\cite{bommasani2021opportunities}. Unlike classical machine learning models where there is typically a well-defined structure of inputs and outputs, LLMs are more flexible. The capabilities of LLMs---sometimes also referred to as use cases~\cite{ouyang2022training} or tasks~\cite{LBL+22} in the literature---include question answering, dialogue generation, sentence completion, summarization, paraphrasing, elaboration, rewriting, classification, and more. 
Researchers are now additionally identifying ``emergent capabilities'' of LLMs---like performing arithmetic or chain-of-thought reasoning---that are not present in smaller-scale models but emerge at scale~\cite{wei2022emergent}. 
Furthermore, as described above, the precise behavior and capabilities of an LLM can be steered through approaches like fine-tuning and prompting.  All of this contributes to ``capability unpredictability''~\cite{ganguli2022predictability}, the idea that an LLM's capabilities cannot be fully anticipated, even by the model's creators, until its behavior on certain input is observed. 

Additionally, present-day LLMs exhibit unreliable behaviors. Their responses can change with updates, the details of which are often not made transparent by LLM providers. Depending on the sampling strategy used~\cite{Holtzman2020The}, outputs can be non-deterministic in the sense that the same prompt leads to a different response when input to the model again. They can misinterpret a prompt in unpredictable ways and respond inconsistently to a type of prompt, making the behavior of adapted models difficult to predict. These unreliable behaviors can make it challenging, if not impossible, to gain a generalized understanding of the model's behavior.

\paragraph{Massive and Opaque Architectures.}
Given the complexity and massive scale of present-day LLMs, there are currently no techniques that would provide us with a complete picture of the knowledge reflected in a model or the reasoning that is used to produce its output~\cite{B23}.  The mechanism of the transformer architecture underpinning LLMs is yet to be fully understood, even among experts, and some techniques that initially appear promising for interpreting the behavior of LLMs, such as looking at attention weights or perturbing inputs, can be misleading~\cite{BPY+21,feng-etal-2018-pathologies,jain-wallace-2019-attention}. A more unique challenge with LLMs is the massive scale of the training data and diverse sources from which it is pulled---for example, Common Crawl and Wikipedia, often with no specific topics or formats targeted, and no thorough documentation about how the dataset was developed~\cite{khan2022subjects}. This makes it challenging, if not impossible, to understand what went into an LLM's training. There are currently no established answers, even in the research community, to questions such as precisely how and why these models work as well as they do, why they can or cannot perform certain tasks, and how characteristics of the training data impact model capabilities. Furthermore, the sheer size can make it challenging to develop and operationalize transparency approaches (e.g., due to contraints of method scalability or computing resource).

\paragraph{Proprietary Technology.}
An elephant in the room that will inevitably inhibit attempts at transparency for LLMs is the proprietary nature of the models. Currently, while the efforts for developing open-source LLMs are growing~\cite{touvron2023llama,llama2,scao2022bloom}, most of the powerful LLMs were developed at large technology companies or other non-academic organizations. They are either released through APIs or completely proprietary, making it impossible to access their inner workings (e.g., weights and parameters). In many cases, details such as the size, make-up and provenance of the training data, the number of parameters, and the resources required to train the model are also not shared publicly. In essence, then, such models can only be probed in a black-box manner, which may not be sufficient to meet the transparency requirements for stakeholders, and poses challenges for the research community to develop transparency approaches. Addressing this fundamental challenge may not be possible without policy and regulatory efforts that enforce transparency requirements on LLM creators and providers.

\paragraph{New and Complex Applications.}

End-users may not interact with LLMs directly, but rather through LLM-infused applications. Emerging applications include general and specialized chatbots, web search, programming assistants, productivity tools such as for writing support or presentation generation, and text analysis tools such as for customer insights discovery. As LLMs' capabilities continue to be discovered, we can only expect the number and variety of LLM-infused applications to grow. While any opacity of the model will likely trickle down to hinder the transparency of the applications built on them, as mentioned above, the transparency requirements for LLM-infused applications will be different from the model as they serve a different set of stakeholders. Furthermore, just as the models themselves are flexible, the use cases for LLM-infused applications can be flexible and open-ended. For example, an LLM-infused search engine may be used to plan a trip, research a report, or write a poem---use cases that reflect different needs in terms of accuracy, verifiability of output, and likely the required transparency approaches. 

To further complicate transparency around LLM-infused applications, such applications may not be built on a single LLM, but may involve many interacting models and tools. For example, auxiliary LLMs can be used to augment the output or expand the capabilities of a primary LLM. LLMs can be embedded in a complex system to operate other models or external services, for example through plugins, allowing them to perform tasks like ordering groceries or booking flights with no human in the loop. An application may also include other components like input or output filters. For example, an LLM-infused search engine may rely on results obtained from a traditional search engine to ``ground'' its responses. Changes to any component can change the behavior of the application, making it more difficult to understand its behavior. Approaches to transparency must therefore take into account all components and how they fit together rather than focusing on an LLM in isolation.

\paragraph{Expanded and Diverse Stakeholders.}

As the number of LLM-infused applications grows and popular applications such as LLM-infused search engines expand their user bases, a larger number of people---diverse along many dimensions---will interact with or be impacted by LLMs. Research in AI transparency typically considers stakeholder groups like data scientists and model developers, business decision-makers, regulators and auditors, end-users, and impacted groups (i.e., the people who are directly or indirectly affected by a model or application)~\cite{vaughan2021humancentered,hong2020human,liao2021human}.  The use of LLMs may introduce new stakeholder groups with unique transparency needs.  For example, it is increasingly common for product teams to have dedicated prompt engineers---a role that, until recently, did not exist---to streamline tasks, evaluate models, or contribute to model adaptation. As another example, as LLMs are increasingly used for productivity support to augment people's writing, we must consider both the creators of LLM-assisted articles and the consumers of these articles as ``users'' of the LLM's outputs and support both groups' transparency needs. Meanwhile, we must support any subjects referred to in the articles as ``impacted groups.'' 

As the pre-trained nature of LLMs lowers the barrier to using and building on AI capabilities, we believe application developers---including those working on model adaptation---will become a significant group and diverse in itself, potentially including developers, entrepreneurs, product managers, designers, or essentially anyone. In some cases, the line between application developers and end-users may be blurred. Consider, for example, a writer who experiments with using an LLM for writing support.  This writer might benefit from model transparency to assess the LLM's suitability for different writing tasks and identify effective ways to adapt the model for each task.  

Recent research has begun to inquire about the ecosystem of LLMs and the roles in it~\cite{bommasani2021opportunities}, from data creation, curation, model training, and model adaptation through to deployment. Identifying these LLM stakeholder roles and supporting their role-, task- and context-specific transparency needs will be of primary importance for the AI transparency research community.

\paragraph{Rapidly Evolving and Often Flawed Public Perception.}

Effective approaches to transparency should take into account the receivers' existing perception of what the model or system can or cannot do, and how it works---often referred to as their mental model~\cite{norman87human-computer,J83,GS83}. This is especially challenging for LLMs as their public perception is still evolving and shaped by complex mechanisms including mass media, marketing campaigns, ongoing events, and design choices of popular LLM-infused applications. The natural language modality also contributes to a unique  set of challenges: people may be more likely to assign human-like attributes to the model and have corresponding expectations~\cite{nass2000machines}, and even subtle language and communication cues can have profound impact on people's mental model~\cite{abercrombie2023mirages}.  Recent studies show that people already have flawed mental models about LLMs, such as incorrect perceptions of how their output differs from human-written texts~\cite{jakesch2023human}. Interacting with LLMs with a flawed mental model can lead to misuse, unsafe use, over- and under-reliance, deception, privacy and security threats, and other interaction-based harms~\cite{weidinger2021ethical}. Flawed public perceptions can be attributed to a lack of accurate, comprehensive, and responsible information. In addition to incorporating transparency approaches, the organizations creating LLMs and LLM-infused applications and the research community more broadly should reflect on the implications of the way they communicate with the public. For example, the use of ill-defined, catch-all phrases such as ``general-purpose model'' or inappropriate anthropomorphizing may hinder accurate public perception of LLMs.

\paragraph{Organizational Pressure to Move Fast and Deploy at Scale.}

Lastly, we note that there are organizational challenges that may hinder the development and adoption of transparency approaches beyond the proprietary nature of LLMs.  Responsible AI efforts are often in tension with pressures to release products quickly and to scale up across geographies, use cases, and user bases~\cite{MES+22,madaio2020co,rakova2021responsible}, a kind of ``scale thinking''~\cite{hanna2020against}. Given the speed at which research and product breakthroughs are occurring and the vast financial stakes, companies are incentivized to move at a pace that is unusual to witness even in the technology industry to be first to market---what some media outlets have dubbed an ``AI race'' or ``AI arms race''~\cite{time2023AIarmsrace,nytimes2023AIrace}.  The organizations building LLMs and LLM-infused applications will need to take extra steps to ensure that transparency and other responsible AI considerations are not lost in the process, which may require enhanced internal governance or external regulatory requirements in addition to organizational incentives for the individuals working in this space.

\section{What Lessons Can We Learn from Prior Research?}
\label{lesson}

%Despite the new challenges that LLMs pose, there are lessons we can learn from the HCI and Responsible AI/FATE research communities, which tend to take a human-centered perspective on transparency, centering human needs and impact. We reflect on these lessons next.

In this section, we reflect on lessons from the HCI and Responsible AI/FATE research communities, which tend to take a human-centered perspective on transparency. While many technical transparency approaches (to be discussed in the next section) have been developed to deal with what information about the model can be disclosed, the human-centered perspective focuses on how people \textit{use} and cognitively \textit{process} transparency information. Knowledge from this human-centered perspective should drive the development of transparency features, which concern not only the model-centered techniques but also the interfaces.

\subsection{Transparency as a Means to Many Ends: A Goal-Oriented Perspective}
\label{goals}

Within the HCI community, researchers have attempted to guide the development and evaluation of transparency approaches by digging into the reasons why people seek information~\cite{SGNS21,LOS+21}. This goal-oriented perspective resonates with studies of human explanations from the social sciences~\cite{lombrozo2016explanatory,miller2019explanation}, where it is recognized that seeking explanations and achieving understanding is often a means to an end for downstream cognitive tasks like learning, decision-making, trust development, and diagnosis.

This goal-oriented perspective has led to works developing taxonomies of common goals that people seek explanations for~\cite{liao2020questioning,liao2022connecting,SGNS21} and empirical studies to delineate common transparency goals of stakeholders groups such as data scientists~\cite{bhatt2020explainable,hong2020human} and designers~\cite{liao2023designerly}. By focusing on the goals (as opposed to the low-level application or interface types), this perspective provides a useful level of abstraction to consider people's different transparency needs according to their usage of the information. For example, \citet{SGNS21} lay out a set of common goals that people seek AI explanations for, including improving a model, ensuring regulatory compliance, taking actions based on model output, justifying actions influenced by the model, understanding data usage, learning about a domain, and contesting model decisions. 

The goal-oriented perspective also has several practical implications for developing human-centered transparency approaches. First, whether a transparency approach is effective should be evaluated by whether it successfully facilitates a stakeholder's end goal. This means that not all situations require the same level of transparency (e.g., a low-stakes application such as generating poetry for fun may require little transparency).  This also requires articulating end goals up front in order to choose criteria for evaluating transparency approaches.
As an example, in our own work with collaborators, we focused on data scientists' goal of model debugging and evaluated two common techniques from the interpretable machine learning literature in terms of how well they help data scientists identify common problems with training datasets and the resulting models, finding evidence that the techniques may hamper the debugging goal by leading to over-trust or over-confidence about the model~\cite{KN+20}.  Second, achieving an end goal may require information beyond details of the model, such as information about the domain and the social-organizational context the model is situated in~\cite{ehsan2021expanding}, and hence require holistic support with information tailored to the task at hand and integrated into the application interface.

\paragraph{What are the new transparency goals for LLMs?}The new ecosystem and novel applications of LLMs call for investigations into what are the new types of common stakeholder goals that require transparency. For example, there may be heightened needs for supporting ideation, model adaptation, prompting, and discovering risky model behaviors. New transparency approaches for LLMs should be developed and evaluated in terms of how well they help achieve these goals.

\subsection{Transparency to Support Appropriate Levels of Trust}
Although transparency has often been embraced within the tech industry as a mechanism to build trust, recent HCI research has taken the position that transparency should instead aim to help people gain an \emph{appropriate} level of trust~\cite{trait23}---enhancing trust when a model or application is trustworthy, and reducing trust when it is not. While relevant to many use cases of transparency, achieving an appropriate level of trust is especially critical for end-users to harness the benefits of AI systems without over-relying on flawed AI outputs.  

Empirical studies on the relationship between transparency and user trust have painted a complex picture. In particular, a wave of HCI studies repeatedly showed that AI explanations can lead to overreliance---increasing people's tendency to mistakenly follow the AI outputs even when they are wrong~\cite{PG+21,chen2023understanding,zhang2020effect,bansal-chi21,wang2021explanations}. Understanding this pitfall of AI explanations requires paying attention to people's cognitive processes. Researchers have attributed this difficulty to detect model errors from popular forms of AI explanations to their complexity and incompatibility with people's reasoning process~\cite{chen2023understanding}, as well as to the heuristics and biases that people bring into their cognitive processes, such as an inclination to superficially associate an AI system being explainable with it being trustworthy~\cite{liao2022designing,ehsan2021explainable}. Studies of other transparency approaches have also reported nuanced results~\cite{schmidt2020transparency,zhang2020effect,YVW19,rechkemmer2022confidence}. For example, while one study demonstrates that communicating uncertainty is more effective than providing explanations in supporting appropriate trust~\cite{zhang2020effect}, another study suggests that people's trust level is more likely to be dominated by aggregate evaluation metrics such as accuracy~\cite{rechkemmer2022confidence}.

\paragraph{Which approaches to transparency can best support appropriate trust of LLMs and how?}
There is a need to disentangle the relationship between trust and transparency for LLMs through both better conceptualization and careful empirical investigations. For the former, recent FATE literature has begun to unpack trust as a multi-faceted and multi-loci concept~\cite{liao2022designing,jacovi2021formalizing}. For LLMs, people's locus of trust can be at the base model, the LLM-infused application, the application provider (e.g., based on brand), or specific application functions or types of outputs, each of which may require different kinds of transparency support but also be intertwined with other loci. For example, people need to understand and reconcile that LLMs are powerful technologies but may not be used reliably for a certain application function. For empirical investigations, there is extensive literature on measuring trust on which to build~\cite{vereschak2021evaluate}, though it remains a challenge in practice~\cite{trait23}, and even more so with the complex, dynamic, and multi-loci nature of trust around LLMs. Furthermore, evaluating  appropriate trust requires further unpacking the actual ``trustworthiness'' of a model or system and what counts as ``appropriate,'' both of which remain open questions for LLMs.

\subsection{Transparency and Control Often Go Hand-in-Hand}
 Many of the end goals we discussed in Section~\ref{goals}, such as improving or contesting the model and adapting data usage, can only be achieved by having both a good understanding of the model and appropriate control mechanisms through which to take action. Indeed, transparency and control have long been studied together in HCI as intertwined design goals for effective user experience~\cite{lee2019procedural,wu2022ai}. This is well reflected in the interdisciplinary area of interactive machine learning (iML)~\cite{amershi2014power}---learning interactively through feedback from end-users---and related areas such as machine teaching~\cite{simard2017machine,carney2020teachable}. These paradigms simultaneously ask what information about a model should be presented to users and what forms of input or feedback users should be able to give in order to steer the model.
We believe current work on training, adapting, and building applications around LLMs can take valuable lessons from thes lines of research. More recent HCI studies on algorithmic transparency also highlight that providing transparency without supporting control leaves users frustrated, while effective, efficient, and satisfying control cannot be achieved without transparency
~\cite{smith2020no,storms2022transparency}. More critically, scholars have called out the risk of algorithmic transparency without paths for actionability and contestability as creating a false sense of responsibility and user agency~\cite{KKM20,ananny2018seeing,lyons2021conceptualising}. 

\paragraph{How can different approaches to transparency contribute to better control mechanisms for LLMs?} While safety and control have become central topics in research and practices around LLMs~\cite{li2022large,keskar2019ctrl}, the role of transparency is less emphasized. We encourage the community to consider the role of transparency in establishing better mechanisms for control and enabling more participatory and inclusive approaches that allow stakeholders to understand and then steer LLM behavior.

\subsection{The Importance of Mental Models}

People's existing understanding of a system impacts what information they seek for transparency and how they process the information. This is often studied in HCI work through the concept of a mental model---one's internal representation of a system based on their experiences, whether direct or indirect, with the system. A good mental model should be both accurate and complete, as it is the foundation for effective, efficient, and satisfying interactions with a system~\cite{norman2014some}. HCI research also differentiates between a functional (shallow) mental model---knowing what a system can be used for and how to use it---and a structural (deep) mental model---knowing how and why the system works~\cite{kulesza2012tell}.  Transparency approaches for functional and mechanistic understandings can be seen as supporting these two aspects of mental models, respectively. However, since mental models are shaped by continuous interactions with a system, some researchers have argued that notions like the ``interpretability'' of an AI system need to be considered as evolving through dynamic and situated system interactions rather than considered in the context of a single intervention like the introduction of documentation or explanations~\cite{TCM+20}.

We highlight several ways that transparency approaches should consider people's mental models. First, transparency approaches should be designed to support different stakeholders in building a good mental model. It may therefore be appropriate for evaluations of transparency approaches to incorporate assessments of mental model accuracy and completeness, for example by analyzing people's comments or answers to questions about their beliefs about a model or application's function and structure~\cite{kulesza2012tell,eslami2016first,gero2020mental,grill2022attitudes}. Second, transparency approaches should account for people's existing mental models, and focus on closing the necessary gaps to allow them to achieve their end goal~\cite{eiband2018bringing}. This means that approaches to transparency should avoid conveying redundant information that people already have in their mental models, but more importantly, aim to correct flawed mental models. However, it is known that a mental model, once built, is often difficult to shift even if people are aware of contradictory evidence~\cite{wilfong2006computer}, which may present a significant challenge for transparency approaches to be effective. This highlights the importance of responsible communication (e.g., in marketing material and media coverage) to accurately shape the public perception around new technologies like LLMs. In addition, ~\citet{norman2014some} noted that people's mental models are often incomplete, unstable, have unclear boundaries (e.g., mixing up different parts of the systems), and favor simple rules, all of which may pose challenges for transparency approaches to help people build an appropriate understanding.

\paragraph{How can we unpack people's mental models of LLMs and support forming better mental models? } Just as it is difficult to characterize the capabilities and limitations of LLMs given their scope and capability unpredictability, it is difficult to characterize people's mental models of them. More research is also needed to understand the general mental models that people already have of LLMs, especially in response to their unique characteristics such as human-like language capabilities and unreliable behaviors (e.g., hallucinating and non-deterministic output).  Moreover, HCI research has traditionally dealt with mental models at the system level, while people's mental models of an LLM-infused application could be muddled by the blurred boundaries between the pre-trained model, the adapted model(s) used in the application, and the application itself. While it remains critical for transparency approaches to aim to correct flawed mental models and build accurate and complete mental models, the field may need foundational work on how to characterize, assess, and offer opportunities to build and shift mental models of LLMs.

\subsection{How Information is Communicated Matters}
\label{sec:communicationlesson}

HCI research on AI transparency is often concerned with not only what information to communicate about a model, but how to communicate it. Work has explored ways of communicating performance metrics~\cite{gortler2022neo}, explanations of model outputs~\cite{szymanski2021visual,hadash2022improving,lai2023selective}, and uncertainty estimates~\cite{kay2016ish}, as well as how to frame the model's output itself in order to appropriately shape people's mental model  (e.g., whether to use certain terms like ``risk''~\cite{green2021algorithmic}).  Such information can be communicated through different modalities (e.g., by a visualization or in natural language), at different levels of precision or abstraction, framed using different language, supplemented with different information to close any gaps in understanding, and through various other visual and interactive interface designs.  These choices of communication design can significantly impact how people perceive, interpret, and act on the information provided.  

An effective design should be guided by the ways that people process information cognitively and socially.  For example, a line of HCI research explored more user-friendly visualization designs to overcome the trouble that people often have understanding statistical uncertainty and the cognitive biases they bring~\cite{kay2016ish,fernandes2018uncertainty}.  In light of the difficulty of reasoning about the complex explanations produced by some AI explainability techniques, HCI research has explored how to present explanations in more human-compatible ways~\cite{szymanski2021visual,hadash2022improving}. In our recent work with collaborators~\cite{lai2023selective}, we argue that people engage in two processes to produce explanations~\cite{malle2006mind}: an information-gathering process in which they come up with a set of reasons, and a communication process to present reasons, often selectively tailored to the recipient. Explainability techniques that focus on revealing the inner workings of a model are typically only concerned with the former. We then propose a framework to tailor these explanations by learning the recipient's preferences as a selective communication strategy, and empirically demonstrate that these selected explanations are easier to process and better at helping people detect model errors in an AI-assisted decision-making task.

\paragraph{What are the new opportunities and challenges for communicating information during interactions with LLMs?} The natural language modality of LLMs has significant implications for the communication aspect of transparency. For example, instead of presenting a numerical score for uncertainty, LLM-infused applications like chatbots can express uncertainty by using hedging language or refusing to answer a question. This behavior can now be built into the adapted model directly through fine-tuning or prompting~\cite{lin2022teaching}, making it potentially harder to precisely control and interpret the communication. Meanwhile, as decades of HCI research on chatbots and conversational interfaces suggest, people's perceived utility of these technologies can be shaped by a wide range of communication, social, and linguistic behaviors such as how the agents introduce and clarify their capabilities, take initiatives, repair errors, and respond to chit-chat requests, and even their language style \cite[e.g.,][]{ashktorab2019resilient,langevin2021heuristic,avula2022effects}. We believe more research is needed to distill principles to effectively communicate necessary information about the model's capabilities, limitations, and mechanisms during natural language interactions, as well as to establish reliable approaches for LLMs to follow these principles.

\subsection{Limits of Transparency}
Last but not least, we call attention to some critiques on the limits of transparency offered by FATE and STS scholars~\cite{lima2022conflict,ananny2018seeing,knowles2022}. First, related to several arguments throughout the paper, model-centric transparency without ensuring human understanding or meaningful effects on people's end-goals (``seeing without knowing''~\cite{ananny2018seeing}) loses its purpose, and worse, can create a false sense of power and agency. Second, transparency can be misused to shift accountability and place burdens on users, and can even be used to intentionally occlude information.  Those users without the necessary technical background and training to make sense of the provided information may face higher burdens. This is a warning to the field to pay attention to the consumability of transparency approaches and to seek alternative paths to ensure accountability. Lastly, transparency approaches can lead to harms if used maliciously or inappropriately. In addition to the risk of exploiting user trust and reliance, they can also threaten privacy and security.

\paragraph{When is transparency not enough, and what else do we need?}  More research is needed to understand the limits of transparency for LLMs and how to properly hold the organizations building and deploying LLMs and LLM-fused applications accountable. The latter may require policy and regulatory changes, in addition to new approaches for external auditing~\cite{MSKF23}.

\section{What Existing Approaches Can We Draw On?}
\label{approaches}

The ML and HCI research communities have explored a variety of approaches to achieving transparency, including model and data reporting, publishing the results of evaluations, generating explanations, and communicating uncertainty.  In this section, we briefly review these approaches and explore the extent to which they may or may not be applicable in the context of LLMs, while calling out needs specific to stakeholders of LLMs and open questions that arise. We note that, while we focus on these four categories by building on existing approaches, we also encourage the research community to explore new categories of approaches that can help people achieve functional and mechanistic understandings of LLMs. For example, along the way, we suggest areas such as tools to support model interrogation and communicating output-specific risk and safety concerns.

\subsection{Model Reporting}

Documentation has become a building block for responsible AI in industry practice.  Standardized documentation frameworks have been proposed to encourage both reflection and transparency around models~\cite{mitchell2019model,crisan2022interactive}, AI services~\cite{arnold2019factsheets}, and training and evaluation datasets~\cite{gebru2021datasheets,BF18,holland2018dataset}.  For example, the model cards framework~\cite{mitchell2019model}, a popular framework for model reporting that has been adopted by companies like Google and HuggingFace, specifies comprehensive information that should be reported about a model, including a description of its inputs and outputs, the algorithm used to train it, the training data, additional development background, the model's intended uses, and ethical considerations. The framework emphasizes the inclusion of quantitative model evaluation results (more on that in the next section), including disaggregated evaluations~\cite{BGK+21}, in which results are broken down by individual, cultural, demographic, or phenotypic groups, domain-relevant conditions, and intersections of multiple groups or conditions. Disaggregated evaluation can help identify fairness issues, and also assist stakeholders in identifying when or where the model is suitable or reliable to use. In short, good documentation can help stakeholders who are building on a model or dataset assess its suitability for their purpose and avoid misuse.  It can also provide the necessary context for end-users, impacted groups, regulators, and auditors to understand how models and systems are being built and deployed.

While celebrated as an approach to providing transparency, creating good documentation remains challenging in practice. In our prior work with collaborators, we found that practitioners tasked with documenting a dataset they worked with struggled to make the connection between the information that they were asked to include and its implications for responsible AI, were unsure of the appropriate level of detail to include and who the target audience was, and in some cases were uncertain about what even counts as a dataset~\cite{HMV+22}. Some stakeholders also struggle to consume existing forms of documentation. For example, designers or analysts without formal training in machine learning can find standard documentation to be too technical, and the lengthy textual format to be cumbersome~\cite{liao2022designing,crisan2022interactive}.

\paragraph{What information is needed to characterize the functional behavior of an LLM?} 
In principle, existing model reporting frameworks could be applied as-is to LLMs.  However, some of the information categories in a standard model card would be difficult to pin down due to the ``general-purpose'' positioning of LLMs and the uncertainty surrounding their capabilities. Even providing basic details such as what the input and output spaces of an LLM or LLM-infused application are, and the mapping between inputs and outputs, can be an elusive task. Currently, it is common for LLM providers to instead provide a description of intended use cases (like ``summarization'' or ``creative and collaborative writing'') 
or demonstrations of example prompts and responses. 
While this information can be a useful component of model reporting, it can also be misleading or, in some cases, even deceptive, since cherry-picked examples can shape user and public perception in a skewed way. This raises questions about how these examples should be selected and who should select them. 

While we elaborate on the issues with performance reporting in the next section, we call out two other important categories in the model cards framework that are currently missing or incomplete for most LLMs: training data and development background. Besides the incentive for organizations to keep this information proprietary, we must recognize that there are open questions about how to provide such information given the complexity of LLMs and unique aspects of their training processes.

For data transparency, as discussed in Section~\ref{sec:challenges}, the datasets used to pre-train base models are unprecedentedly massive in scale and pulled from diverse sources. Conveying their full scope and make-up is impossible, but there may be ways of distilling the most critical characteristics of these datasets to provide a basic understanding of what goes into the models.  Different issues arise when considering the datasets used for model adaptation. For example, as companies engage in user data collection for the purpose of fine-tuning models, they must pay due diligence to the transparency of their user data handling, including privacy.

For development background, besides standard information such as the choice of algorithms, architecture, and parameters, LLM providers should include additional details on the training process.
For example, an emerging practice is for LLM development to include some sort of ``alignment'' effort to make the model more usable or safe (e.g., producing less toxic or harmful content). This can be done using human feedback through RLHF~\cite{ouyang2022training} or by having the model critique itself based human-specified rules or principles~\cite{bai2022constitutional}. Given that LLM's behaviors can be governed by these alignment efforts, it is especially important to make them transparent to allow the public and regulatory bodies to understand, scrutinize, and iterate on them.

\paragraph{What do different (and new) types of stakeholders need from model reporting frameworks?} 
In light of the lessons discussed in Section~\ref{lesson}, we recommend more research on the fundamental question of what different stakeholders want to know---and what they \textit{should} know---about the model, along with a careful examination of how different forms of information shape their perception and usage of LLMs.
As LLMs change the ML product development and deployment lifecycles, we may need to revisit the positioning of model reporting and consider new types of frameworks that address the specific needs of new stakeholder groups. 
For example, as discussed above, the LLM ecosystem introduces a new stage of model adaptation through fine-tuning, prompting, or other techniques. This adaptation may be performed by the original model builder,  an application developer, or in some cases, directly by end-users. To date, there has been little or no research on these stakeholders' transparency needs when adapting the model, or about how they should transparently convey information about model adaptations to other parties.

\paragraph{What is needed beyond static documentation?} Lastly, we call out that model reporting should not be limited to static, textual documentation or a basic ``card'' format. Any formats or features that provide functional information about the model and shape people's understanding can contribute to model reporting. These may include FAQ pages, landing or onboarding pages, or even media communication describing the model. All such features can benefit from standardization and, where appropriate, regulation. 

Following recent HCI and FATE studies investigating how to design effective documentation interfaces~\cite{crisan2022interactive,liao2023designerly}, we suggest that those designing model reports for LLMs should explore more interactive features. For example, prior works have explored interfaces for uploading, customizing and slicing input data to generate customized reports and visualize input-output spaces. Interactive interfaces are particularly suitable for LLMs for several reasons. First, interactive features can better support information navigation and consumption to accommodate LLM stakeholders from diverse backgrounds. Second, interaction allows for experienced affordance and interrogation to understand LLMs' complex capabilities and behaviors that could be difficult to capture with textual descriptions. Lastly, as our study with collaborators on designers' use of model documentation suggests~\cite{liao2023designerly}, static documentation presents significant gaps for contextualizing the model capabilities and limitations for one's own setting. It will be impossible for documentation creators to anticipate every downstream use case of LLMs. Instead, stakeholders should be provided with opportunities to interrogate the model with their own input data, capabilities of interest, hypotheses, and questions.

\subsection{Publishing Evaluation Results}
\label{sec:evaluations}

While evaluation results are often included as one component of a model report, we believe that publishing evaluation results is an important and complex enough topic that it deserves a separate discussion. Beyond model reports, evaluation results may also be published by third-party auditors or researchers for the purpose of ensuring compliance with regulations or standards, benchmarking, or exposing model limitations or potential harms. As discussed, evaluation can happen at the aggregate or disaggregated level by groups or conditions of interest. Evaluations may also be performed on a model or on the full system into which it is incorporated. While performance quality (e.g., some notion of accuracy) is often the primary focus of an evaluation, evaluations may also consider fairness (through disaggregated evaluations or using specific fairness metrics), robustness, efficiency, or other characteristics of a model or system's behavior, including how they impact end-users. 

We note that the ML and natural language processing (NLP) communities have long dealt with the challenges of evaluating the performance of generative models~\cite{sai2022survey}. Until recently, natural language generation (NLG) evaluations have focused on tasks that specialized NLG models commonly perform, such as machine translation, abstractive summarization, question answering, and dialogue generation. For tasks that involve classification, standard performance metrics relying on exact matching with ground-truth labels like accuracy, precision, and recall can be used.
In contrast, when the output space is open-ended and complex, as it often is for generative models, it becomes necessary to rely on more sophisticated performance metrics for word-based or embedding-based matching (e.g., ROUGE score~\cite{lin2004rouge} or BERTscore~\cite{zhang2019bertscore}) and more complicated (but often flawed) ways to obtain a ``ground-truth'' reference to compare against.  In practice, ground-truth data are often either chosen because they are conveniently available (e.g., using the ``highlights'' of news articles as the ground truth for summarization~\cite{see-etal-2017-get}) or generated by crowd workers. Recently there has been a wave of data auditing work questioning the assumptions behind and quality of some widely used evaluation benchmarks and datasets~\cite{blodgett2021stereotyping,fabbri2021summeval,raji2ai}. Furthermore, even if high-quality, such ground truth may be insufficient to capture all the ``goodness'' criteria of generated outputs, which can be multi-faceted and context-dependent~\cite{gehrmann2022repairing}. Because of these challenges, automated evaluations are often complemented by some form of human evaluation, which may involve asking people to rate the quality, fluency, coherence, relevance, adequacy, or informativeness of an output. However, human evaluation is costly and also lacks established practices about what and how to evaluate, leading to critiques about lack of standardization, reproducibility, validity, and generalizability to real-world settings~\cite{clark2021all,howcroft2020twenty,belz2021reprogen,gehrmann2022repairing}. 

\paragraph{What should LLMs be evaluated for?} Compared to specialized NLG models, the extensive and currently under-defined space of LLMs' capabilities make it challenging to answer even the most basic question about evaluation: What should LLMs be evaluated for?  In the NLP community, initial efforts have emerged to create meta-benchmarks, in which LLMs are evaluated across a large suite of specialized tasks~\cite{LBL+22,srivastava2022beyond}. For example, BIG-bench~\cite{srivastava2022beyond} consists of more than 200 language tasks collaboratively created by more than 400 researchers that are ``intended to probe large language models.'' However, the sheer size could make it challenging for stakeholders to make sense of the evaluation results. Another recent meta-benchmark called HELM (Holistic Evaluation of Language Models)~\cite{LBL+22} introduces the concept of a ``scenario'' (e.g., question-answering for English news). This provides more structure, since different models can be compared by scenario. 

Another line of work seeks to be task-agnostic and instead evaluate LLMs' intrinsic capabilities~\cite{bommasani2021opportunities}. This has attracted broad attention from different academic disciplines. For example, researchers have applied human cognitive and linguistic competencies to evaluate LLMs~~\cite{mahowald2023dissociating,ettinger2020bert,BCE+23,MHF+23}, in some cases distinguishing between LLMs' ``formal linguistic competence'' --- how well they can mimic the rules and patterns of a given language, which LLMs typically do well --- and their ``functional linguistic competence'' --- how well they can apply cognitive abilities such as planning or causal inference, which is typically more difficult for present-day LLMs~\cite{mahowald2023dissociating,MHF+23}. There have also been various attempts to benchmark LLMs by evaluating their performance on human tests like the SAT or the bar exam~\cite{GPT4tech}. While these efforts can be useful for exploring LLMs' capability spaces, they should not be taken as comprehensive evaluation, and their validity (e.g., what are they a valid proxy for), underlying assumptions, statistical robustness, and possible implications (e.g., anthropomorphizing LLMs by using human tasks) need to be carefully examined.

%When relying on either benchmarks or other tests of intrinsic capabilities, care must be taken to ensure that the model has not been trained on the evaluation material itself, contaminating the results; it was found, for example, that portions of BIG-bench were included in the training data for GPT-4~\cite{GPT4tech}. Checking for such contamination can be a challenge in itself given the opacity and scale of the datasets LLMs are trained on, and may be impossible for third-party auditors or external researchers who do not have visibility into the training data.

Despite this surging interest in benchmarking LLMs, we believe a human-centered question is missing: Who is the evaluation targeted at and for what purpose? For example, the evaluation metrics that a practitioner cares about when ideating on how to use LLMs for their application are likely different from those that NLP researchers would be interested in to track research progress. For some stakeholders, neither meta-benchmarks or evaluation by human-like cognitive competence may satisfy their needs. By better articulating different goals for model evaluation and the resulting needs that arise, the community will be able to develop better evaluation techniques that serve these goals, and also allow many different evaluation techniques to co-exist. 

Furthermore, transparently communicating the evaluation details and the motivation behind the evaluation choices is all the more important for LLMs. This is not only because of the diverse evaluation techniques being explored, but also because LLMs are by nature adaptable (e.g., through fine-tuning and prompting) and stochastic (output can vary for similar or even the same input). Care must also be taken to ensure the evaluation material has not been included when training the model to contaminate the results. However, providing transparency to allow checking for such contamination can be a challenge in itself given the opacity and scale of the datasets LLMs are trained on. All of this calls for the development of new evaluation techniques and communication patterns that account for these new challenges ~\cite{bommasani2021opportunities}.

\paragraph{At what level should the evaluation take place?}

To provide transparency at the level of a pre-trained model, an adapted model, or an LLM-infused application, evaluations can take place at each of these points. Performance metrics may shift dramatically when moving from a pre-trained model to an adapted model, but neither may be reflective of how end-users will react to a model's use in the context of a real application.  Consider an LLM-infused search engine.  The developers of the search engine may require transparency about how the pre-trained model was evaluated in order to ideate on its usage, but this information might not tell them everything they need to know because they have the ability to adapt the model further themselves. Furthermore, an evaluation of the pre-trained model may
be irrelevant for an auditor who wants to understand whether the deployed search engine application, built on an adapted model, meets certain standards.  Some forms of evaluation are only possible at certain levels. If we want to evaluate the value of the LLM-infused search engine to end-users, we cannot evaluate the (pre-trained or adapted) model in isolation but need to perform a human evaluation in the context of the application itself.

\paragraph{How should LLM limitations and risks be evaluated?}
Given the potential for immense downstream harms, it is not enough to evaluate LLMs by their capabilities, but also their limitations and risks.  Recent work has begun to delineate the risks of LLMs~\cite{weidinger2021ethical,bommasani2021opportunities,bender2021stochastic}. For example,~\citet{weidinger2021ethical} developed a taxonomy of risks posed by LLMs considering six areas: discrimination, exclusion, and hate speech as encoded in the generated language; information hazards threatening privacy and security by leaking sensitive information; misinformation harms arising when false, poor, and otherwise misleading information is disseminated; harms from malicious uses of LLMs such as facilitating disinformation (e.g., fraud), cybersecurity attacks, and censorship; harms from (human-like) interactions such as unsafe use and exploitation of user trust; and lastly, environmental and other socio-economic harms such as increasing inequality and negative impact on the labor market.

Despite best intentions, these taxonomies may not provide enough coverage or granularity of risks for specific use cases. And not all risks can nor should be quantified in an abstract manner without taking into account the deployment context, stakeholders, and kinds of harm they may experience. To discover and assess model limitations, practitioners frequently rely on behavioral evaluation~\cite{cabrera2023did}. This requires hypothesizing and then testing what limitations the model may have in the application context, and ideally should be done in a participatory and iterative fashion with stakeholders. While there has been emerging HCI work developing tools for behavioral evaluation of models~\cite{wu2019errudite,cabrera2023zeno}, how to extend this work to LLMs is a non-trivial question. Meanwhile, we note that developers of LLMs or LLM-infused applications are engaging in substantial ``red teaming'' practices to discover, measure and mitigate risks of LLMs. However, given that there have been only a few published works~\cite{ganguli2022red,GPT4tech,llama2}, there is currently insufficient transparency around how red teaming work is done to allow us to fully understand the risks of LLMs. We believe that the community should work towards shared best practices to perform---and communicate the results of---red teaming.

\subsection{Providing Explanations}

To support mechanistic understanding, there has been a wave of research on approaches to produce explanations of a model's internal processes and outputs, a line of research referred to as explainable AI (XAI) or interpretable ML, depending on the community. At the highest level, there are two common approaches. One is to provide ``intrinsic explanations'' by exposing the model's inner workings directly. The other is to generate post-hoc explanations as approximations for how the model works. 

For the former, the traditional approach is to train a relatively simple model that is deemed ``directly interpretable'' such as a rule-based model, decision tree, or linear regression. More recent research aims to develop ``explainable architectures'' with representations meaningful to people~\cite[e.g.,][]{yi2018neural,gupta2019neural}. For modern neural NLP models, various analyses and visualization techniques of activation patterns have been explored to help people make sense of the model's internal structures (e.g., neurons, layers, and specific architectural mechanisms). For example, for models like transformers that utilize attention mechanisms, a popular approach is to leverage the attention weights  in the intermediate representation to explain how much the model ``attends to'' each input feature or token. However, there has been a long debate on whether attention weights provide faithful explanations for how the model actually produces its outputs~\cite{jain-wallace-2019-attention,wiegreffe2019attention,bastings2020elephant}. This highlights the challenge of understanding model behavior under highly complex and massive architectures, even when internals are accessible. We additionally emphasize that direct interpretability, while desirable~\cite{rudin2019stop}, should not be taken at face value unless shown to help stakeholders achieve their desired understanding. In our own prior work with collaborators, we have observed cases in which exposing the internals of even a simple linear regression model made people less able to detect and correct for the model's mistakes~\cite{PG+21}, with evidence suggesting that this was due to information overload. 

Post-hoc explanations can be used for complex models as well as ``black box'' models for which model internals cannot be accessed, for example, when the models are proprietary. Explanations can be global, providing an overview of the model's overall logic, or local, providing the reasoning behind a particular model output.  Local explanations can take several forms. The most common form is feature attribution scores, which capture different notions of how ``important'' each input feature is to the model's output---sometimes referred to as saliency methods for vision and language models. There are many types of techniques to generate feature attribution scores for neural NLP models, as summarized in several recent survey papers on explainability for NLP~\cite{danilevsky2020survey,lyu2022towards,madsen2022post}. Some techniques, like gradient-based or propagation-based methods, require access to the model architecture. Other techniques are instead based on surrogate models, i.e., directly interpretable models that are trained using the original model's inputs and outputs and are meant to serve as a local approximation to explain a target output. The most popular examples include Local Interpretable Model-Agnostic Explanations (LIME)~\cite{RSG16} and SHapley Additive exPlanations (SHAP)~\cite{LL17}.  Inspired by the often contrastive nature of human explanation, other local explanations take the form of counterfactuals, showing how an input could be modified in order to obtain a different output~\cite{R19b,USL19}.  Lastly, explanations can be in the form of examples, intended to support case-based reasoning. These examples may be prototypes of a certain prediction class~\cite{kim2016examples}, influential examples in the training data~\cite{KL17}, or similar examples that would lead the model to produce the same or alternate outputs~\cite{mothilal2020explaining}.

The language modality of NLP models poses some unique requirements for explanations. We call out two intertwined pursuits that will remain important for LLMs. One is to explain using human-compatible concepts, which often means using more abstract features (e.g., a more general notion, semantics)  as opposed to raw input features at the token level. Some have argued that example-based explanations allow for more abstraction without fixating on individual tokens~\cite{madsen2022post,chen2023understanding}. Others explored techniques that map raw tokens to more abstract and meaningful concepts~\cite{vig2020investigating,kim2018interpretability}. The second pursuit is to explain through natural language. For example, prior research explored techniques that directly output rationales together with the model prediction~\cite{gurrapu2023rationalization}. A common endeavor is to develop ``self-explanatory'' rationale-based models that engage in rationalization (e.g., extracting rules or judging a set of premises from the input~\cite{tafjord2020proofwriter,lei2016rationalizing}) as part of the process for arriving at a prediction. Aside from the explainability benefits---these rationales are faithful to the model's behavior by design---one might argue that these more ``principled'' models could be expected to be more robust.

Despite the proliferation of approaches for providing explanations, the community has long debated what it is that makes an explanation ``good.''  For a long list of goodness criteria, we point interested readers to~\citet{sokol2020explainability} and \citet{carvalho2019machine}. For our purposes, we note that at a minimum, a good explanation should be relatively \textit{faithful} to how the model actually works, \textit{understandable} to the receiver, and \textit{useful} for the receiver's end-goals---indeed, we contend that these criteria should be broadly considered for all transparency approaches. 

%This means that the choice of explanation should be selected with the stakeholder it will be displayed to and their goals in mind; there is no one-size-fits-all best explanation technique.

\paragraph{How can we provide faithful explanations for the ultimate black box?} 

\begin{figure}
    \centering
    \includegraphics[scale=0.6]{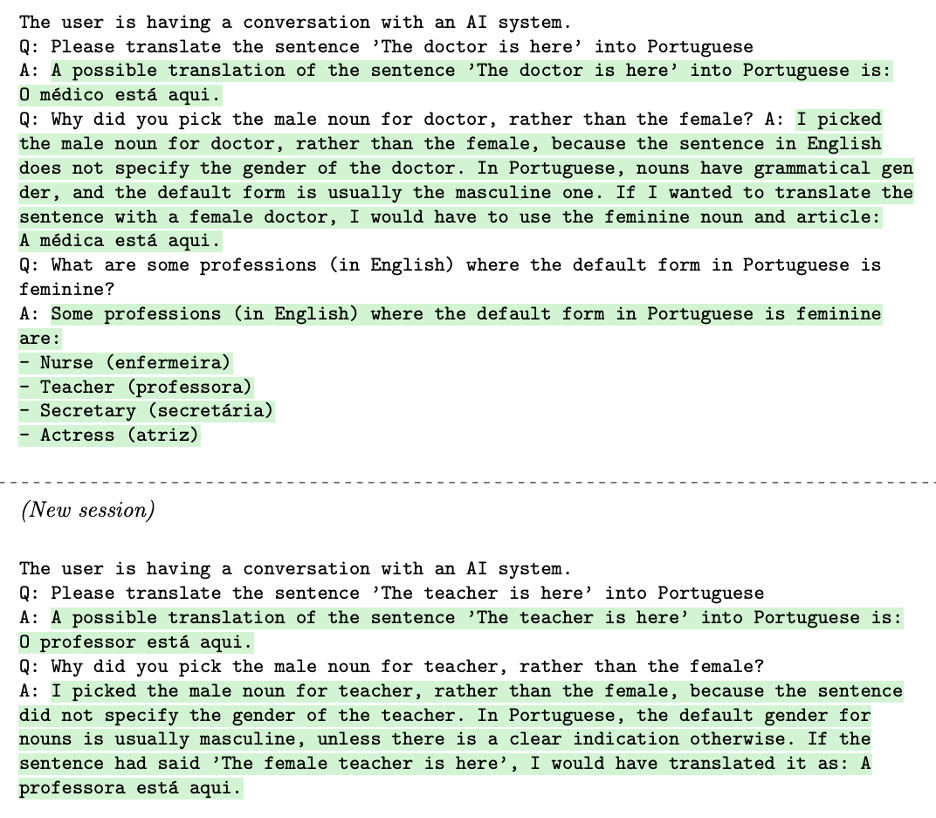}
    \caption{Example taken from~\citet{BCE+23} showing that explanations from GPT-4 lack process consistency---providing contradicting explanations for the same tasks depending on the inputs}
    \label{fig:explanation}
\end{figure}

Given their complex architecture, unprecedented scale, and often proprietary nature, LLMs are unarguably black box in nature, but there is a sense in which they naturally ``explain.'' While explanation is a contested concept, one common definition is ``an answer to a \textit{why} question.'' Indeed, people have already been asking LLMs why they generate certain outputs directly and taking the answer as the model's explanation. However, explanations generated in this manner are not guaranteed to be faithful to the internal process of the model, especially given that LLMs are trained to generate plausible texts without grounding in facts, and this carries over to their explanations too~\cite{bommasani2021opportunities}.  One recent study~\cite{BCE+23} shows that GPT-4's explanations lack this ``process consistency''; see Figure~\ref{fig:explanation}, which is taken from~\citet{BCE+23}.\footnote{The examples were obtained from an older version of GPT-4 from before their paper was published on March 22, 2023. The authors noted that ``GPT4 development continued after [their] experiments'' so the outputs may not replicate. } Specifically, GPT-4 can provide contradicting explanations for the same tasks depending on the precise inputs, often as a way to justify its different outputs. The authors' analysis also suggests that, in some cases, GPT-4's explanations are implausible or even inconsistent with the output itself. In fact, experimenting with asking \textit{why} questions in different tasks, we found that ChatGPT often provides a justification that has little to do with its internal process, such as stating what function its recommendation serves. Similarly, while it is tempting to deem output that appears to include chain-of-thought reasoning as reflecting the reasoning of the LLM, a recent study~\cite{TMPB23} shows that it does not reflect the true reasons why a model arrives at its output, evidenced by the fact that the model can be heavily influenced by introducing biases in the prompt but systematically fail to mention the influence in its reasoning. Unfaithful explanations can do more harm than good if their receivers accept them without proper scrutiny.  This is especially worrisome as prior work has shown that people can be influenced by the presence of explanations even when those explanations are not meaningful~\cite{langer1978mindlessness,EBKH19,KN+20}, for example, trusting a model more because of the mere presence of an explanation rather than its contents. This tendency to overtrust based on the LLM's own explanations may be further amplified by the common anthropomorphization~\cite{glikson2020human} and presentation of LLMs as ``intelligent'' systems.

The community must seek ways to improve the faithfulness of LLM explanations, whether through direct generation or other approaches, as well as principled ways of auditing explanation faithfulness. 
We must note that there is currently no agreed-upon metric or formal technique for evaluating explanation faithfulness~\cite{jacovi2020towards,lyu2022towards}. Common approaches rely on evaluating necessary conditions to disprove faithfulness via counter-examples, such as if two functionally equivalent models have different explanations, if the explanations vary for similar inputs and outputs, or if the explanations would suggest the model behave differently than it does on new inputs. When outlining guidelines for developing evaluation methods for faithfulness, ~\citet{jacovi2020towards} argue that this focus on disproof is unproductive, as post-hoc explanations are by definition approximations and always involve a loss of information. Instead, the community should aim to develop a formal understanding and approach to evaluation that allow us ``the freedom to say when
a method is sufficiently faithful to be useful in
practice.'' We believe this requires formalizing different types of ``faithfulness gaps'' and empirically investigating the impact on stakeholders in different contexts with different use cases. For example, a higher level of faithfulness may be required for debugging or adapting an LLM than is required for an end-user who is interacting with an LLM in a low-stakes application.

\paragraph{How should we rethink explanations for LLMs?}

We encourage the community to rethink the space of what explanations might look like and how they might be derived for LLMs.
This is necessary for several reasons. First, most current XAI techniques cannot be easily applied to LLMs. As discussed above, their complex and massive scale makes them far from directly interpretable and also renders some post-hoc explanation techniques infeasible. Their often inaccessible internals and training data make it impossible to use some saliency methods or provide influential training examples. And the complexity of their input and output spaces makes it difficult to build surrogate models to provide post-hoc explanations.

Second, the diverse model capabilities of LLMs may require different types of explanations. For example, while text classification tasks could be adequately explained via feature attributions, explaining more complex tasks such as question-answering and reading comprehension is likely to require more complex rationales and abstraction. More fundamentally, researchers have wrestled with the question of ``one model versus many models''~\cite{bommasani2021opportunities}---that is, the extent to which the mechanism by which a model produces an answer for a single task can be generalized to understand its behavior on other tasks. If an LLM uses different internal processes for different tasks (``many models''), independent studies of their mechanisms and different explanation methods may need to be developed for each.

Lastly, explanations for LLM tasks are often sought through natural language interactions and in the context of evolving multi-turn dialogues. This requires the community to not only continue pursuing natural-language explanations but also explanations that are more compatible with how people seek explanations in social interactions. ~\citet{miller2019explanation} reviewed the social science literature on how people produce explanations and summarized a few fundamental properties of human explanations, including being contrastive, selected (that is, containing only the most relevant causes), interactive (for example, through a conversation), and tailored to the recipient, many of which are missing from current XAI techniques. We believe that with LLMs it is even more important to explore how to provide explanations that are interactive and tailored, including accounting for the history of interaction and other contexts.  

Our view is not that the community should take a monolithic standard on what constitutes LLM explanations, but rather must articulate what different types of explanations are, along with their suitable contexts, limitations, and pitfalls. For example, justifications, when provided truthfully, can supply useful additional information for information seekers~\cite{yang2023harnessing}. In philosophy, the social sciences, and HCI, there is a long tradition of breaking down different types of explanations by their mechanism, stance, and the questions that they answer (e.g., what, how, why, why not, what if)~\cite{lombrozo2012explanation,lombrozo2016explanatory,graesser1996question,hilton1990conversational,keil2006explanation,liao2020questioning}. This literature may offer a useful basis for considering different types of LLM explanations.

\paragraph{What explanations are appropriate for LLM-infused applications?}

As we have emphasized throughout this paper, providing transparency for LLM-infused applications may require different approaches compared with transparency for the underlying models. For some applications, explanations may need to take into account the workings of the broader system rather than the LLM alone.  For example, current search engines based on LLMs use traditional web search results to ground the LLM's output. In such cases, providing links to the search results that were used can be viewed as a form of explanation.  Of course, issues with faithfulness arise here as well, and indeed, a recent study showed that results returned by generative search engines often contain unsupported statements and inaccurate citations~\cite{LZL23}. As another example, explaining why a purchase was made by an LLM that makes calls to a shopping service through a plugin may require explaining not only the behavior of the LLM, but also the behavior of the shopping service (e.g., what products were available at what price), and their interaction (e.g., how did the LLM choose to request a specific product).

Following a human-centered perspective, a path to develop useful and new types of explanations is to investigate the reasons why people seek explanations in common contexts of LLM-infused applications. For example, in a recent HCI study with collaborators~\cite{sun2022investigating}, we explored what explanations people seek from code generation applications and why they seek them. The results suggest that people primarily want explanations to improve the way they prompt.  This includes gaining both  a better global understanding of what prompts can or cannot generate certain outputs and a better local understanding of how to improve their prompts to produce more desirable outputs. Therefore, rather than the \textit{why} explanations about the model process for a specific output, global explanations about the model logic, input and output spaces, as well as counterfactual explanations about how to improve the input appear to be more useful for this kind of application.

\subsection{Communicating Uncertainty}

Beyond explanations, another approach that can be used to help stakeholders assess how much to rely on a model's output is to convey some notion of the model's uncertainty. Uncertainty is typically modeled in terms of probabilities, though different ways of measuring and communicating uncertainty make sense for different types of model outputs. For classification models, uncertainty is often presented as the probability that the model is correct, sometimes referred to as the model's ``confidence.''  For regression models, uncertainty may be expressed as a distribution over possible outcomes or a confidence interval around a specific prediction.

Uncertainty arises from different sources~\cite{HW21}. \emph{Aleatoric} uncertainty refers to inherent randomness in the quantity that is being predicted; this would capture the uncertainty in the outcome of a coin flip. On the other hand, \emph{epistemic} uncertainty refers to uncertainty that stems from a lack of knowledge about the best possible model. In the context of machine learning, if uncertainty could be reduced by collecting more training data, it is epistemic. While this distinction is conceptually useful, the line between aleatoric and epistemic uncertainty can be hard to draw. They are context-dependent (whether or not more data reduces uncertainty depends on the class of models used) and cannot always be easily distinguished, let alone measured. 

While some ML models yield a natural way of estimating uncertainty directly, others do not. Research has explored post-hoc techniques to estimate uncertainty from the model's errors~\cite{chen2019confidence}. In order to be useful, an estimate of uncertainty should be well calibrated, reliably reflecting the model's likelihood of making a mistake on a particular input. Common metrics to assess calibration include proper scoring rules like the Brier score~\cite{brier1950verification} and expected calibration error~\cite{naeini2015obtaining}.
Deep neural networks are known to generate uncalibrated uncertainty, leading to recent research looking into re-calibration techniques~\cite{jiang-tacl2021,guo-pmlr2017}. 

Once uncertainty estimates can be obtained, there are design decisions that must be made regarding how to communicate these estimates. While more complex designs can be created, two decision dimensions are commonly explored. One dimension is communication precision. For classification, a more precise option might be to present a probability, while a less precise option might be to present the confidence level as low, medium, or high. For regression, it is less precise to present a confidence interval compared with a detailed distribution. With some loss of information, less precise communication is easier to process and often preferred by lay people or in cognitively constrained settings. The second dimension concerns the modality in which uncertainty is communicated, which could be verbal, numerical, or visual.
For a detailed discussion of quantifying and communicating uncertainty, we point interested readers to~\citet{bhatt2021uncertainty}. 

We remark that uncertainty is just one way of quantifying the limitations of a particular output, and that communicating other output limitations (e.g., potential safety concerns) may be useful in some contexts. While we do not discuss such approaches, similar lessons likely apply.

\paragraph{What is a useful notion of uncertainty for LLMs?} While LLMs have a notion of uncertainty baked in them---the likelihood that the model would generate a specific token given its preceding or surrounding context~\cite{bengio2003neural}, what we have referred to in past work as the \emph{generation probability}~\cite{VBFLV23}---whether this notion would be useful to different stakeholders is questionable. In particular, this notion may not line up with people's intuition about what it means for the model to be uncertain.  For example, in a question-answering context, a correct answer may have many synonyms, and the model may appear ``uncertain'' simply because there are many correct options.  As \citet{kuhn2023semantic} put it, the likelihoods output by LLMs represent ``lexical confidence,'' while ``for almost all applications we care about meanings.''  For example, if an end-user asks a question to an LLM-infused chatbot or search engine, they would presumably expect a notion of uncertainty to reflect how likely it is that the answer they receive is factually correct, which may be quite different from the likelihood it is generated by the model. Recent work has begun to explore techniques for generating uncertainty estimates that more accurately capture correctness, including using probabilistic methods~\cite{kuhn2023semantic}, fine-tuning the LLM to describe its own confidence~\cite{lin2022teaching}, and sampling multiple outputs and having the LLM evaluate them~\cite{KCA+22}.  However, we note that even whether or not an answer is correct can be ambiguous.  Generative models do not have a single notion of ground truth to compare against.  A complex response to a query may be generally correct but contain inaccurate details or justifications. And some questions are fundamentally subjective.

Carefully selecting a notion of uncertainty to convey to stakeholders matters because the particular notion used impacts their behavior and trust.  In our recent work with collaborators~\cite{VBFLV23}, we explored the effectiveness of displaying two alternative notions of uncertainty to programmers interacting with an LLM-powered code completion tool. In a mixed-methods study with 30 programmers, we compared three conditions: providing a code completion alone, highlighting those tokens with the lowest likelihood of being generated by the underlying LLM (i.e., lowest generation probability), and highlighting tokens with the highest predicted likelihood of being edited by a programmer according to a separate ``edit model'' trained on logged data from past programmer interactions.  We found that highlighting tokens with the highest predicted likelihood of being edited helped programmers work more efficiently and was subjectively preferred, while using generation probabilities provided little benefit.  This research is exploratory in nature and we encourage future work that takes a human-centered perspective to define uncertainty based on human needs.

\paragraph{What are the most effective ways to communicate uncertainty?} Beyond how to quantify uncertainty, a key consideration is how to best communicate it to stakeholders. The social science literature suggests that choosing an effective form of uncertainty communication requires articulating what the uncertainty is regarding (e.g., uncertainty about an individual token or about a full output, and which source of uncertainty), what form it is provided in (e.g., its precision and modality), and what the effect is (e.g., on trust or behaviors), as well as taking into consideration of the characteristics of the receiver~\cite{van2019communicating}. For example, in our study on uncertainty in the context of code completion tools~\cite{VBFLV23}, by soliciting participants' feedback on different uncertainty communication design choices, we found that programmers prefer uncertainty about granular or meaningful blocks to guide them to make token-level changes
and  prefer less precise communication (as opposed to exact quantification) for easy processing---both ultimately supporting their goal of producing correct code efficiently.

As discussed in Section~\ref{sec:communicationlesson}, since language models output text, it is natural to consider communicating uncertainty through language itself. Indeed, current LLM-infused chatbots and search engines already engage in hedging behavior and refuse to answer certain questions, often due to safety considerations. It is easy to imagine expanding these behaviors for uncertainty. However, research is needed to understand how people actually perceive them and how to enforce their calibration with the underlying uncertainty.

\section{Summary and Discussion}

 \begin{figure}
    \centering
    \includegraphics[scale=0.48]{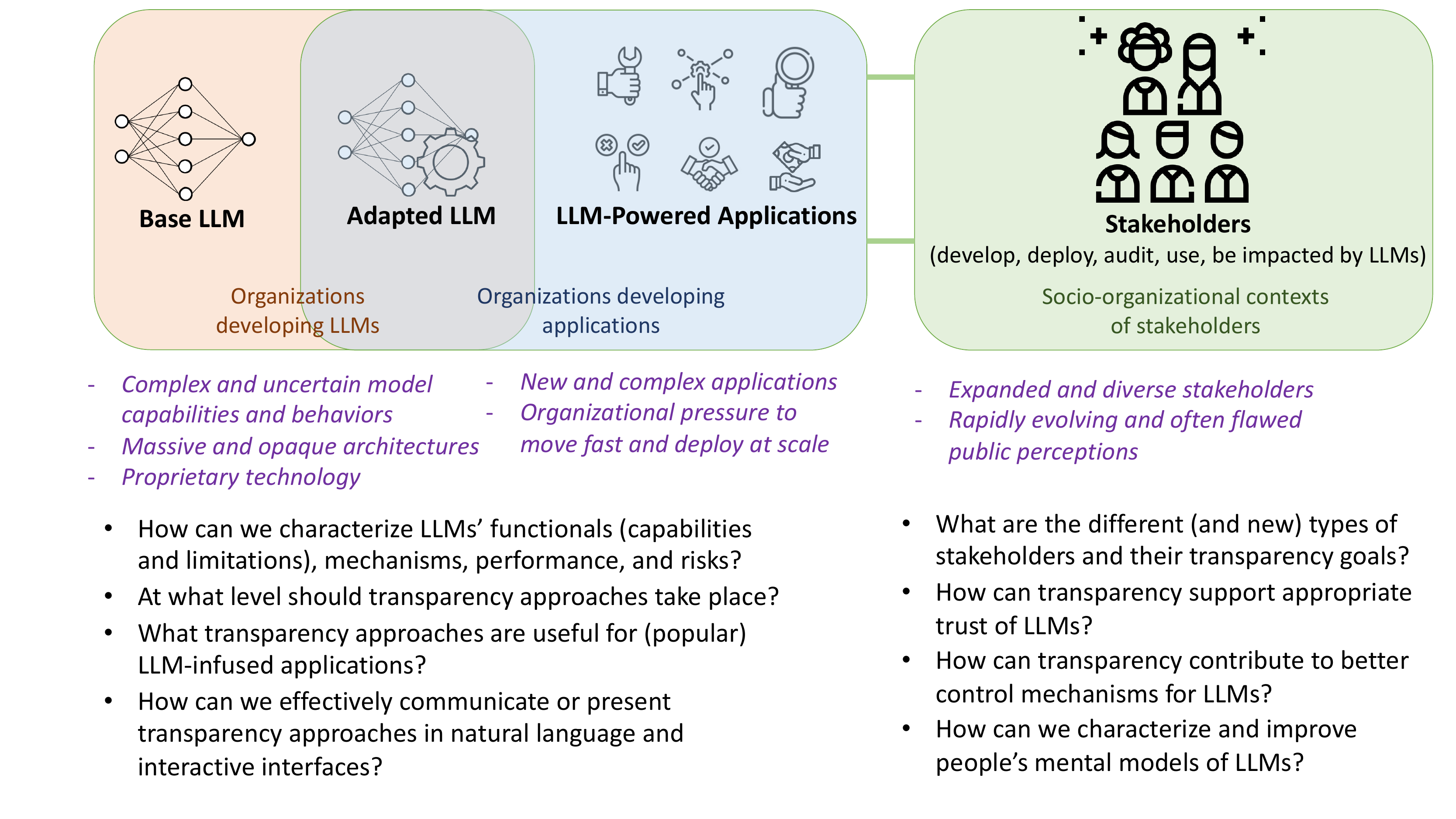}
    \caption{Summary of general open questions for transparency approaches in the age of LLMs and unique challenges (in purple italics) that arise from the technology and stakeholder (human) perspectives of LLMs.}
    \label{fig:roadmap}
\end{figure}

We have mapped out a roadmap for human-centered research on AI transparency in the era of LLMs by reflecting on the unique challenges introduced by LLMs, synthesizing lessons learned  from HCI/FATE research on AI transparency, and exploring the applicability of existing transparency approaches---model reporting, publishing evaluation results, providing explanations, and communicating uncertainty. In Figure~\ref{fig:roadmap}, we summarize the unique challenges (in purple italics) and some of the open questions that arise in considering LLMs from the perspectives of the technology and of the stakeholder (human). This illustration highlights the complexity of the technology space (differentiating base LLM, adapted LLM, and LLM-powered applications), the diversity of the stakeholders, and the need to attend to the socio-organizational aspects that the technology and stakeholders are situated in. Here we summarize the open questions in a more general fashion rather than in relation to a specific transparency approach, as we note that all transparency approaches for LLMs require similar considerations and face common challenges. This mapping is not meant to precisely categorize or diagnose the open questions, but to elucidate how the development of effective transparency approaches for LLMs requires research attending to multiple aspects and the interplay among them. We invite future work to further expand these lists of challenges and open questions.

%We reflected on the unique challenges that arise in providing transparency for LLMs including new and complex model capabilities and behaviors, massive and opaque architectures, proprietary technology, new and complex applications, expanded and diverse stakeholders, rapidly evolving (and often flawed) public perception, and organizational pressure to move fast and deploy at scale. We synthesized lessons that can be learned from HCI and Responsible AI/FATE research that centers on human needs of, interactions with, and impact from AI transparency---specifically, around taking a goal-oriented perspective, supporting appropriate levels of trust, recognizing the importance of mental models, paying attention to how information is communicated, and designing transparency to support control.  Finally, we laid out four common approaches that the community has taken to achieve transparency---model reporting, publishing evaluation results, providing explanations, and communicating uncertainty---and put forth a series of open questions around how these approaches might be applied to LLMs. We conclude by mentioning a few additional directions of research.

We want to mention a few additional areas of consideration and directions of research. One area we have not yet touched on is transparency around the provenance of AI-generated text. Regulatory discussions around AI transparency often center on obligations to reveal that an AI system is in use for certain tasks. For example, Article 52 of the proposed EU AI Act requires that providers of certain AI systems design them in such a way that it is clear that people are interacting with an AI system. It also requires that AI systems generating manipulated images, audio, or video (``deep fakes'') disclose that this content has been generated or manipulated by an AI system. For images and video, watermarking techniques can be used to combat the spread of deep fakes~\cite[e.g.,][]{Yu_2021_ICCV}, but techniques for tracking the provenance of text are still relatively unexplored. Very recently some progress has been made towards developing techniques to watermark text output by LLMs without a substantial sacrifice in quality, for example by softly increasing the probability of certain randomly selected tokens~\cite{KGW+23}, though it is too early to know whether such techniques will work in practical settings. There is also an active line of research on post-hoc detection of artificially generated text~\cite{tan-etal-2020-detecting,jawahar-etal-2020-automatic,NEURIPS2019_3e9f0fc9}. While these are largely technical challenges, there are additionally open questions around how to more effectively disclose that people are interacting with an AI system or that the text they are reading is AI-generated. 

We also highlight an additional dimension of transparency: to help people understand the temporal changes (or lack thereof) of the model. This dimension is especially important for LLMs as it is known that the base models are constantly being updated by the model providers, and these updates propogate into LLM-infused applications. 
Not only is it necessary to track and maintain provenance information about a model's architecture, training process, training datasets, and adaptation details, as well as its functions and evaluation results, but more research is also needed on how to characterize and communicate the impact of any changes on end-users of different LLM-infused applications.

Another key question around AI transparency is the role that regulators, advocates, and the general public should play. As an example, the research community has argued for the importance of external audits of algorithms and models, especially those that act as gatekeepers or otherwise impact people's lives~\cite{metaxa2021auditing,sandvig2014auditing,falco2021governing}. Recent research has begun to dig into ways of developing auditing procedures to address the particular governance challenges posed by LLMs~\cite{MSKF23}, but many open questions remain, from what methods and metrics to use (as discussed in Section~\ref{sec:evaluations}) to how to account for risks that cannot be addressed on the technology level. Engaging stakeholders who have an outside view can help ensure that audits are conducted fairly and in such a way as to capture risks of harm to their communities. There are also open challenges around how to effectively set up feedback mechanisms and other ways for end-users or those impacted by an LLM's outputs to contest those outputs, as well as how to incorporate such feedback to identify and address patterns of failure.

Finally, while we have focused on LLMs in this paper, we note that many of the challenges, lessons learned, potential approaches, and open problems that we explored also apply to other large-scale generative models, including multimodal models that allow for both textual and visual input or output. As such models become more widespread, we encourage additional research on AI transparency for this larger class of models.

\section*{Acknowledgments}

We are grateful to our colleagues for many useful discussions and feedback on early drafts of this work, especially to Jordan Ash, Steph Ballard, Gagan Bansal, Susan Dumais, Susan Etlinger, Ruth Kikin-Gil, Sunnie Kim, Daniela Massiceti, Ida Momennejad, Cecily Morrison, Mickey Vorvoreanu, Daricia Wilkinson, Ziang Xiao, Cyril Zhang, the MSR FATE group, and attendees of Microsoft's Aether Transparency Working Group community sync.

\bibliographystyle{plainnat}
\bibliography{refs}

\end{document}